\documentclass[aps,pra,superscriptaddress,twocolumn,showpacs]{revtex4-1}
\usepackage{dcolumn}
\usepackage[dvips]{graphicx}
\begin{document}
\title{Correlated spontaneous symmetry breaking induced by zero-point
fluctuations\\ in a quantum mixture}
\author{Li Ge}
\affiliation{State Key Laboratory of Surface Physics, Department of Physics, Fudan University, Shanghai 200433, China}
\author{Fei Zhou}
\affiliation{ Department of Physics and Astronomy,  University of British Columbia, Vancouver, Canada, V6T 1Z1}
\author{Yu Shi}
\email{yushi@fudan.edu.cn.}
\affiliation{State Key Laboratory of Surface Physics, Department of Physics, Fudan University, Shanghai 200433, China}

\begin{abstract}
We propose a form of spontaneous symmetry breaking driven by zero-point quantum fluctuations. To be specific, we consider the low-energy dynamics of a mixture of two species of spin-$1$ Bose gases. It is demonstrated that the quantum fluctuations lift a  degeneracy regarding the relative
orientations of the spin directors of the two species, and
result in correlation or locking between these  macroscopic variables. This locking persists in the presence of the trapping potential and weak magnetic fields,
allowing, in principle, an experimental probe of this correlated spontaneous symmetry breaking, as a  macroscopic manifestation of zero-point quantum fluctuations.
\end{abstract}

Phys. Rev. A {\bf 89}, 043623 (2014)


\maketitle

\section{Introduction}

Spontaneous symmetry breaking is a central concept in all of physics~\cite{anderson}. The
quantum collective phenomena induced by zero-point motion of dynamical variables are also
ubiquitous~\cite{lamb,anderson,shender,henley,coleman} and
the dynamics governed by the
quantum fluctuations of the collective degrees of freedom are extremely important for our
understanding of macroscopic emergent  phenomena~\cite{caldeira}.
In this paper we demonstrate a different type of  spontaneous symmetry breaking, namely, one in which the
macroscopic variables of two interacting many-body systems are locked together through
microscopic zero-point quantum fluctuations.  We frame our theory in the context of spinor Bose
gases~\cite{law,koashi,ho,Stamper12}, where quantum fluctuations can be dominating in energetics or dynamics and are highly controllable~\cite{zhou2,turner,barnett,song08,cui}.  Specifically, we consider  a  mixture of two
spinor Bose gases with
interspecies spin exchanges~\cite{shi1,luo,xu,shi2} and show that in the symmetry-breaking ground state,
the spin directors of the
two gases are locked together by the zero-point quantum fluctuations. The spin director  is the spin  direction modulo $Z_2$ symmetry due to the compensation of the inversion of the spin direction  by a $\pi$ phase transformation~\cite{zhou1}. Our results give more motivation to study interspecies spin exchanges in mixtures of different spinor Bose gases~\cite{shi1,luo,xu,shi2},  which are under experimental investigation~\cite{xiong}.

This article is organized as follows.
In Sec.~\ref{hm} we define the many-body Hamiltonian, and discuss the mean-field theory. In Sec.~\ref{qf} we consider quantum fluctuations and  find  the energies in the Bogoliubov theory. In Sec.~\ref{lock} we investigate the fluctuation-induced locking between the directors of the two species. The fluctuation-induced spin dynamics is discussed in Sec.~\ref{spind}.  In Sec.~\ref{trap} we discuss how the locking effect survives a trapping potential or an external magnetic field. Finally, we summarize our investigation in Sec.~\ref{summ}.

\section{Hamiltonian and Mean Field Theory \label{hm} }

The many-body Hamiltonian we apply to study this problem is
\begin{equation}
\mathcal{H}= \sum_{\alpha=a,b}\mathcal{H}_\alpha+\mathcal{H}_{ab},
\end{equation}
$$\begin{array}{rl}
{\cal H}_{\alpha} &=  \int d\mathbf{r}
\psi^{\dagger}_{\alpha\mu}h_{\alpha} (\mathbf{r})_{\mu\nu}
\psi_{\alpha\nu} \\ &
+ \frac{1}{2} \int d\mathbf{r}
\psi^{\dagger}_{\alpha\mu}\psi^{\dagger}_{\alpha\rho}
({c}_0^{\alpha}\delta_{\mu\nu}
\delta_{\rho\sigma}+{c}_2^{\alpha}
\mathbf{F}_{\alpha\mu\nu}\cdot\mathbf{F}_{\alpha\rho\sigma})
\psi_{\alpha\sigma}\psi_{\alpha\nu}
\end{array}  $$
is the Hamiltonian of species $\alpha$. Here
$h_{\alpha}=-\frac{\hbar^2}{2m_{\alpha}}\nabla^2+V_{\alpha} (\mathbf{r})$ is the spin-independent
single-particle Hamiltonian of each atom of species $\alpha$, $c_0^\alpha$ is the intraspecies density-density interaction strength of
species $\alpha$, $c_2^\alpha$ is the intraspecies spin-exchange interaction strength of
species  $\alpha$, and $\psi_{\alpha\mu}$ represents  the field
operator for the $\mu$ component of species $\alpha$, with  $\alpha=a,b$ and  $\mu = -1, 0, 1$ or $x, y ,z$,  depending on the basis used. For the case of $\mu=x, y, z$, the $\eta$ component $F^{\eta}_{\mu\nu}$ of
$\bf{F}_{\mu\nu}$ is $- i \epsilon_{\eta \mu\nu}$, where
$\epsilon_{\eta\mu\nu}$ is the  Levi-Civit\`{a} antisymmetric tensor~\cite{Snoek04}. In addition,
$${\cal H}_{ab}= \int d\mathbf{r}
\psi^{\dagger}_{a\mu}\psi^{\dagger}_{b\rho}
({c}_0^{ab}\delta_{\mu\nu} \delta_{\rho\sigma}+{c}_2^{ab}
\mathbf{F}_{a\mu\nu}\cdot\mathbf{F}_{b\rho\sigma})
\psi_{b\sigma}\psi_{a\nu}$$
is the interaction between the two species, with  $c_0^{ab}$ the interspecies
density-density interaction strength and  $c_2^{ab}$ the interspecies spin-exchange
interaction strength.  Repeated indices are summed over.  We focus on the regime of  $c_2^a>0$,
$c_2^b>0$ and $c_2^{ab}>0$ and   assume that $c_2^ac_2^b>(c_2^{ab})^2$.

The three-component vector field ${\mathbf \psi}_\alpha$ of the species $\alpha$
can be written as
\begin{equation}
\label{nematic} {\mathbf \psi}_\alpha = \Phi_\alpha\mathbf{n}_\alpha,
\end{equation}
where $\mathbf{n}_\alpha$ is the spin director , which is also a three-component vector, and  $$\Phi_\alpha \equiv \sqrt{\rho_\alpha }e^{i\chi_\alpha},$$ where
$$\rho_\alpha=\psi^{\dagger}_{\alpha\mu}\psi_{\alpha\mu}$$ is the number-density operator.  In terms of
$\rho_\alpha$ and the  spin-density operator $$\textbf{L}_\alpha=
\psi^{\dagger}_{\alpha\mu}\bf{F}_{\alpha\mu\nu}\psi_{\alpha\nu}, $$ the Hamiltonian can be
rewritten as
\begin{equation}
\mathcal{H}=\mathcal{H}_p+\mathcal{H}_s,
\end{equation}
where  $$ \begin{array}{rl}
\mathcal{H}_p=&\sum_{\alpha=a,b}\int d\mathbf{r}[\frac{1}{2m_\alpha}|\nabla
\Phi_\alpha(\mathbf{r})|^2+V_a(\mathbf{r}) \rho_\alpha(\mathbf{r})  \\ &+
\frac{1}{2}c_0^\alpha\rho_\alpha^2(\mathbf{r})]  +\int
d\mathbf{r}[c_0^{ab}\rho_a(\mathbf{r})\rho_b(\mathbf{r})] \end{array} $$ is the phase  part,
$$ \begin{array}{rl} \mathcal{H}_s = & \sum_{\alpha=a,b}\frac{1}{2}\int
d\mathbf{r}[\frac{\rho_\alpha (\mathbf{r})| \nabla \mathbf{n}_\alpha(\mathbf{r})|^2}{m_\alpha}
+c_2^\alpha\mathbf{L}_\alpha^2(\mathbf{r})]\\ & +\int
d\mathbf{r}[c_2^{ab}\mathbf{L}_a(\mathbf{r}) \cdot\mathbf{L}_b(\mathbf{r})]  \end{array}  $$  is  the spin part, and a spin-phase coupling part is negligible in the long-wavelength limit or when
$\mathbf{L}_\alpha =0$.
Hence the phase  and spin degrees of freedom are decoupled and are described by the collective
variables $\{\rho_\alpha(\textbf{r}),\chi_\alpha(\textbf{r})\}$ and
$\{\textbf{n}_\alpha(\textbf{r}), \textbf{L}_\alpha(\textbf{r})\}$,  respectively.  Here  $\mathcal{H}_p$ simply describes a
mixture of two scalar Bose gases, and is independent of the relative orientation of
$\textbf{n}_a$ and  $\textbf{n}_b$. Henceforth  we focus on the spin part $\mathcal{H}_s$.

First consider the uniform case  $V_a=V_b=0$. In the ground state, $\rho_a$ and $\rho_b$ are
both constants and the total energy
\begin{eqnarray} 
E =& {\cal V} (
\frac{1}{2}c_2^a
\mathbf{L}_a^2+\frac{1}{2}c_2^b\mathbf{L}_b^2+
c_2^{ab}\mathbf{L}_a\cdot\mathbf{L}_b \nonumber  \\
& +
 \frac{1}{2}c_0^a \rho_a^2+\frac{1}{2}c_0^b\rho_b^2+ c_0^{ab}\rho_a\rho_b), \end{eqnarray}  
where ${\cal V}$ is the volume of the system. Minimization of $E$ implies that in the ground
state, $\Phi_a$, $\Phi_b$, $\textbf{n}_a$ and $\textbf{n}_b$ are
all position independent, while $\textbf{L}_a=\textbf{L}_b=0$, implying that the total spin is also $0$.
The mean-field ground state is $|N_a,\mathbf{n}_a\rangle \otimes
|N_b,\mathbf{n}_b\rangle$, where each species is in its own spin nematic state uncorrelated with the other species. Here $|N_\alpha,\mathbf{n}_\alpha\rangle$
denotes the state in which the spin director of each atom of species $\alpha$ is aligned along the direction of
$\mathbf{n}_\alpha$.
There are no constraints on $\textbf{n}_a$ and $\textbf{n}_b$, hence the ground state
manifold becomes $\frac{S^1\times S^2}{Z_2}\otimes \frac{S^1\times S^2}{Z_2}$, which possesses
a huge degeneracy.

\section{Quantum Fluctuations \label{qf}}

\begin{figure}
\begin{center}
\scalebox{0.6}{\includegraphics[-20pt,5pt][523pt,260pt]{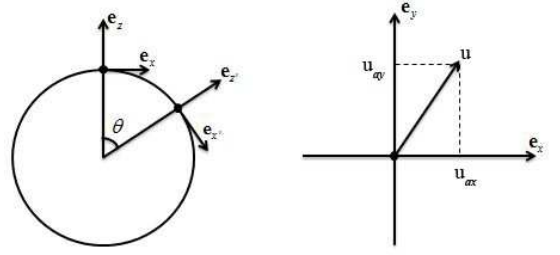}}
\caption{ \label{fig1} The   diagram on the left shows a typical mean-field configuration of the two nematic vectors   $\textbf{n}_a^0=\textbf{e}_z$ and $\textbf{n}_b^0=\textbf{e}_{z'}$, with an arbitray angle $\theta$. The   diagram  on the right shows the small fluctuation $\mathbf{u}_z$ around  $\textbf{n}_a^0$. The fluctuation $\mathbf{u}_b$  of $\textbf{n}_b$ in the $x'y'z'$ frame is similar. The coordinate systems are set such that  $\textbf{e}_y=\textbf{e}_{y'}$.  }
\end{center}
\end{figure}

In the following  we show that this degeneracy is lifted by zero-point  quantum fluctuations. Denoted by  $\textbf{n}_a^0$ and $\textbf{n}_b^0$,  the spin directors of the two species in  a mean-field symmetry-breaking ground state satisfy the relation  $\textbf{n}_a^0 \cdot  \textbf{n}_b^0 =
\cos\theta $.  Let us arbitrarily set $\textbf{n}_a^0 =\textbf{e}_z$, $\textbf{n}_b^0 =\textbf{e}_{z'}$ and $\textbf{e}_z\cdot \textbf{e}_{z'}=\cos \theta$, as depicted in  Fig~\ref{fig1}.     The  degeneracy implies that $\theta$ is arbitrary. The range of $\theta$ is
limited to $-\pi/2 \leq \theta < \pi/2$ because of $Z_2$ symmetry.

Consider
\begin{equation}
\textbf{n}_\alpha = \textbf{n}_{\alpha}^0+\textbf{u}_\alpha,
\end{equation}
for $\alpha =a,b$, where   quantum fluctuations are
$$ \textbf{u}_a \simeq\frac{\psi_{ax}}{\sqrt{\rho_a}}\textbf{e}_x+
\frac{\psi_{ay}}{\sqrt{\rho_a}}\textbf{e}_y,$$
$$\textbf{u}_b\simeq
\frac{\psi_{bx}}{\sqrt{\rho_b}}\textbf{e}_{x'}+
\frac{\psi_{by}}{\sqrt{\rho_b}}\textbf{e}_{y},$$ with  unit vectors  $\textbf{e}_x$, $\textbf{e}_y$, and
$\textbf{n}_a^0$ forming a Cartesian coordinate system with  $\textbf{e}_{x'}$, $\textbf{e}_{y}$,
and $\textbf{n}_b^0 $ forming another one, as shown in Fig.~\ref{fig1}.
We find that
\begin{equation}
\mathcal{H}_s= \sum_{i=x,y} \mathcal{H}_{si},
\end{equation}
where 
\begin{eqnarray}
\mathcal{H}_{si}=&\sum_{\alpha,\mathbf{k}}\frac{k^2}{2m_\alpha}
\psi_{\alpha i,\mathbf{k}}^{\dagger}\psi_{\alpha i,\mathbf{k} }+\frac{1}{2}c_2^\alpha\rho_\alpha\sum_{\alpha,\mathbf{k}}[2\psi_{\alpha i,\mathbf{k}}^{\dagger}\psi_{\alpha i,\mathbf{k} }\nonumber \\ &-(\psi_{\alpha i,\mathbf{k} }^{\dagger}\psi_{\alpha i, -\mathbf{k}}^{\dagger}+H.c.)]
 + c_2^{ab}\sqrt{\rho_a\rho_b} \zeta_i(\theta) \nonumber \\
&\sum_{\mathbf{k}}(\psi_{a i,\mathbf{k} }^{\dagger}\psi_{b i,\mathbf{k} }-\psi_{a i,\mathbf{k} }^{\dagger}\psi_{b i,-\mathbf{k} }^{\dagger}+H.c.),
\end{eqnarray}
written in terms of momentum $\mathbf{k}$, with  $k\equiv |\mathbf{k}|$, $\zeta_x(\theta) =\cos\theta$, and  $\zeta_y(\theta)=1$. Here
$\mathcal{H}_{si}$ depends only on $\psi_{\alpha i}$,   $\mathcal{H}_{sx}$  depends on $\theta$,  and $\mathcal{H}_{sy}$ is independent of $\theta$.

By performing a Bogoliubov transformation,
one obtains
\begin{equation}
\label{diag}
\mathcal{H}_{s i}=\sum_{\lambda=\pm,k} \omega_{i\lambda, k} (A_{i\lambda,k}^{\dag}A_{i\lambda,k}+\frac{1}{2}),
\end{equation}
where $A_{i\lambda,k}$ is some  Bosonic operator and
\begin{eqnarray}
\omega_{i\pm,k}^2= & \frac{1}{2}
\big( \epsilon_{ak}^2+\epsilon_{bk}^2 \pm [(\epsilon_{ak}^2+\epsilon_{bk}^2)^2
-4E_{ak}E_{bk}(4g_ag_b  \nonumber \\ & -4g_{ab}^2\zeta^2(\theta)+2g_aE_{bk}+2g_bE_{ak}+
E_{ak}E_{bk})]^{1/2} \big),  \nonumber \\
\end{eqnarray}
with $E_{\alpha k} \equiv \frac{k^2}{2m_\alpha}$, $\epsilon_{\alpha k}^2
=E_{\alpha k}^2+2g_\alpha E_{\alpha k}$, $
g_\alpha \equiv c_2^\alpha\rho_\alpha$,  and $g_{ab} \equiv c_2^{ab}\sqrt{\rho_a\rho_b}$.
Note that $\omega_{x\pm,k}$ depends on both $g_{ab}$ and $\theta$, while $\omega_{y\pm,k}$ depends on $g_{ab}$ but is independent of  $\theta$.

\section{Fluctuation-induced Locking \label{lock}}

According to the spectra of ${\cal H}_{sx}$ and ${\cal H}_{sy}$  obtained above, we know that the quantum fluctuations lead to  a $\theta$-dependent zero-point energy
\begin{equation}
{\cal E}_{0}(\theta)=E_0(\theta)+ E_0(\theta=0),  \label{e00}
\end{equation}
where on the righthand side, the first term $E_0(\theta)$
is the zero-point energy of ${\cal H}_{sx}$, with
\begin{equation}
E_0(\theta)=\frac{1}{2} \sum_k  \Omega_{k}(\theta),
\end{equation}
where
\begin{eqnarray}
\Omega_{k}(\theta) &\equiv&
\omega_{x+,k}+\omega_{x-,k} \nonumber \\
&= &
\big[\epsilon_{ak}^2+\epsilon_{bk}^2+2E_{ak}^{1/2}E_{bk}^{1/2}
(4g_ag_b-4g_{ab}^2\cos^2\theta \nonumber \\
&&+2g_a E_{bk}+2g_b E_{ak}
+E_{ak}E_{bk})^{1/2}\big]^{1/2},
\end{eqnarray}
which reaches its minimum at $\theta=0$.  The second term $E_0(\theta=0)$ on the righthand side of (\ref{e00}), which is $\theta$-independent,  is the zero-point energy of ${\cal H}_{sy}$.

Hence the total zero-point energy ${\cal E}_{0}(\theta)$ also reaches its minimum at $\theta=0$.
Therefore, in the ground state,  $\textbf{n}_a^0$ and $\textbf{n}_b^0$  are actually locked in the low-energy limit, that is,  they tend to align in the same direction. This is in contrast to what was suggested by the mean-field analysis.

For large $k$, the $\theta$-dependent part of $\Omega_{k}$  behaves as
$$-\frac{ g_{ab}^2 m_am_b}{k^2(m_a+m_b)}\cos^2\theta+O(\frac{g_{ab}^4}{k^6}), $$ thus the $\theta$-dependent part of the energy has ultraviolet divergence. This divergence originates from the use of contact interaction, which fails at short range or large momentum. It can be removed by introducing
a momentum cutoff or by the renormalization of interaction strengths, that is, by evaluating  the ground-state energy in terms of the renormalized
quantities $c^{ab}_{0,r}$ and $c^{ab}_{2,r}$,  which are directly related to the experimentally observed scattering lengths and correspond to the bare quantities $c^{ab}_0$ and $c^{ab}_2$ respectively.

It is known that $c_{0}^{ab}=\frac{2U_2+U_0}{3}$ and $c_{2}^{ab}=\frac{U_2-U_0}{3}$, where $U_0$ and $U_2$ are the interaction strengths
for the total spin $F=0$ and $2$ channels, respectively~\cite{luo}. Accordingly $c_{0,r}^{ab}=(2U_{2,r}+U_{0,r})/3$ and
$c_{2,r}^{ab}=(U_{2,r}-U_{0,r})/3$, with $$U_ {F,r}=\lim_{k\rightarrow 0} \langle \mathbf{k}',F |\hat{T}| \mathbf{k},F\rangle,$$ with $k=|\mathbf{k}|= |\mathbf{k}'|$,     given by the zero-energy $\hat{T}$-matrix element for two-body scattering~\cite{ueda}.  The
Lippman-Schwinger equation  reads  $$\hat{T}=\hat{U_F}+\hat{U_F}G_0\hat{T},$$  with $\hat{U_F}=U_F\delta(\mathbf{r})$ is the two-body potential and  $$G_0(k)=-\frac{2M}{k^2},$$ where  $$M \equiv \frac{m_am_b}{m_a+m_b}, $$ is the zero-energy Green's function  for the relative motion of atoms $a$ and $b$. Consequently,  $\frac{1}{U_{F,r}}=\frac{1}{U_{F}}+\int d^3k\frac{2M}{k^2}$. Since the divergence
occurs in the second order of interaction strengths, we expand the above formula to second order and obtain $$U_F= U_{F,r}+U^2_{F,r}\int d^3k\frac{2M }{k^2}.$$
Therefore,  \begin{eqnarray}
c_0^{ab}&=&\frac{2U_2+U_0}{3}\nonumber \\ &=& \frac{2U_{2,r}+U_{0,r}}{3} +\frac{2U^2_{2,r}+U^2_{0,r}}{3}\int d^3k\frac{2M}{k^2} \nonumber \\
        &=& c_{0,r}^{ab}+[(c_{0,r}^{ab})^2+2(c_{2,r}^{ab})^2]\int d^3k\frac{2M }{k^2}, \label{c0}
        \end{eqnarray}
which is essential for the cancellation of the divergence.

Renormalization effect should be considered for all the terms of $c_0^{ab}$ and $c_2^{ab}$,   including those in the mean field energy, in fluctuations of ${\cal H}_p$ and ${\cal H}_s$. In  the mean field energy,  by substituting  (\ref{c0}) for  $c_0^{ab}\rho_a\rho_b$  in the mean field energy, it can be seen that it becomes  $c_{0,r}^{ab}\rho_a\rho_b$ after the $(c_{0,r}^{ab})^2$ term cancels the
divergent term in the Bogliubov ground state energy of $H_p$, while $(c_{2,r}^{ab})^2$ term cancels the divergent  $(c_2^{ab})^2$ terms in ${\cal E}_0(\theta)$, with exact cancellation at $\theta =0$.

Therefore, the  zero-point energy ${\cal E}_{0}(\theta)$ should be regularized by using
\begin{equation}
E_0(\theta)=\frac{{\cal V}}{2} \int \frac{d^3 k}{(2\pi)^3}[ \Omega_{k}(\theta) -\frac{\partial \Omega_{k}(\theta) }{\partial g_{ab}^2}g_{ab}^2 ],
\end{equation}
where the summation has been replaced with an integral.
It can be shown that after the
substraction, $\frac{\partial E_0}{\partial\theta}|_{\theta=0}=0$ and  $\frac{\partial^2
E_0}{\partial\theta^2}|_{\theta=0}>0$ still hold, thus $E_0(\theta)$ remains  minimal at
$\theta=0$.

Without a loss of the generality, we focus on  the case $g_a= g_b= g$.  By introducing
$k=k_0x$, where  $k_0 \equiv \sqrt{2gM}$ is  a characteristic momentum,   we rewrite
$E_0(\theta) ={\cal V} k_0^3g I (\frac{g_{ab}}{g},\cos^2\theta)$, where
$$I (\frac{g_{ab}}{g},\cos^2\theta) \equiv \int \frac{d^3 x}{(2
\pi)^3}f\Big(\frac{g_{ab}}{g},\cos^2\theta,x\Big)$$ is dimensionless and $f$ is also dimensionless and depends on $g_{ab}/g$, $\cos^2\theta$, and the dimensionless quantity $x$. Hence
\begin{equation}
E_0(\theta)= g \sqrt{2M^3} N_\alpha \sqrt{\rho_\alpha (c_2^\alpha)^3} I
(\frac{g_{ab}}{g},\cos^2\theta),
\end{equation}
where $\sqrt{\rho_\alpha (c_2^\alpha)^3}$ is analogous to the Lee-Huang-Yang parameter in
dilute gas theory~\cite{lhy}.

\begin{figure}
\begin{center}
\includegraphics[width=10cm]{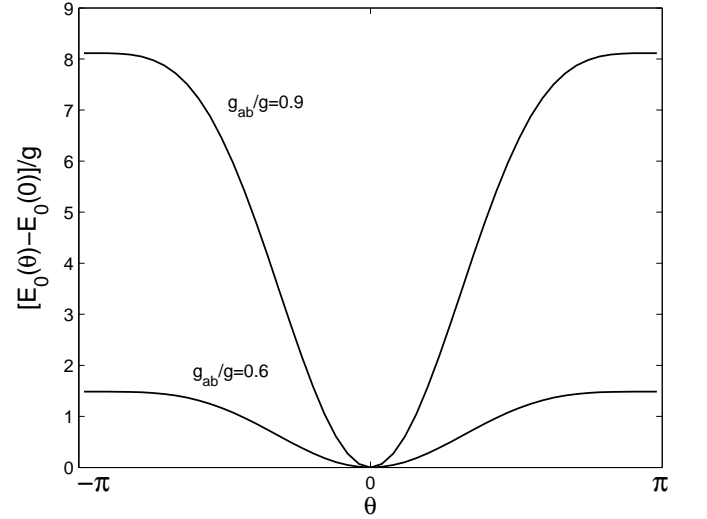}
\caption{ \label{fig2} Zero-point energy $E_0(\theta)$    as a function of the  angle $\theta$ between the spin directors $\mathbf{n}_a$ and $\mathbf{n}_b$. For simplicity,  the dimensionless quantity $[E_0(\theta)-E_0(0)]/g$ is shown as the vertical coordinate. Note that $\theta$ is equivalent to $\theta+\pi$ because of the $Z_2$ symmetry. Here $g_{ab} \equiv
c_2^{ab}\sqrt{\rho_a\rho_b}$ and
$g_\alpha \equiv c_2^\alpha\rho_\alpha$, with  $c_2^{\alpha}=\frac{4 \pi \hbar^2 \Delta
a_a}{m_a} $,   $\Delta a_\alpha$  being  the difference between the triplet and singlet
scattering lengths for atoms of species $\alpha$.  In addition,  $g_b=g_a=g$ is assumed without loss of generality.   The parameter values  are set  as  $\Delta a_a=0.1$nm,  $\rho_a=10^{15}{\rm cm}^{-3}$,
$m_b/m_a=3.3$, and  $N_a=10^4$.  }
\end{center}
\end{figure}

The behavior of $E_0(\theta)$  for various values of $g_{ab}/g$ was  numerically investigated and is shown in Fig.~\ref{fig2}.
The result indicates that  $E_0(\theta)$
increases with $g_{ab}$ because of the enhancement of spin fluctuations.
The zero-point energy plays the role of an effective potential and can strongly
influence the coherent spin dynamics.
Although the mean-field spin dynamics has already been studied in quite a few
laboratories~\cite{Schmaljohann04,Chang05,Higbie05,Stamper12,Widera05}, the macroscopic quantum spin dynamics driven
by microscopic quantum fluctuations so far has not yet been explored in
experiments and remains to be observed. The phenomenon studied here provides   motivation and a different venue in which to understand fluctuation-induced dynamics.

\section{Fluctuation-induced spin dynamics \label{spind}}

In the present case, the effective Hamiltonian that controls the fluctuations of the spin
directors is
\begin{equation} \label{homo}
\mathcal{H}_{eff}=\sum_{\alpha }\frac{c_2^\alpha}{2\Omega}\mathbf{l}_{\alpha
}^2+\frac{c_2^{ab}}{\Omega }\mathbf{l}_{a}\cdot \mathbf{l}_{b}+E_0(\theta)+E_0(0).
\end{equation}
where $\textbf{l}_\alpha \equiv \Omega\textbf{L}_\alpha$.  Defining the center-of-mass
quantities
$\mathbf{l}=\mathbf{l}_a+\mathbf{l}_b$  and
$$
\mathbf{n}=\frac{(c_2^b-c_2^{ab})\textbf{n}_{a}+(c_2^a-c_2^{ab})
\textbf{n}_{b}}{c_2^a+c_2^b-2c_2^{ab}},$$
 and the  relative quantities
$$\mathbf{l}_r=\frac{(c_2^a-c_2^{ab})\textbf{l}_{a}-(c_2^b-c_2^{ab})
\textbf{l}_{b}}{c_2^a+c_2^b-2c_2^{ab}}$$ and
$\mathbf{n}_r=\mathbf{n}_a-\mathbf{n}_b$, we can rewrite $\mathcal{H}_{eff}$ as
$\mathcal{H}_{eff} = \mathcal{H}_{c}+\mathcal{H}_{r}$, where
$$\mathcal{H}_{c}=\frac{1}{2\Omega}\frac{c_2^ac_2^b-(c_2^{ab})^2}{c_2^a+c_2^b-2c_2^{ab}}
\mathbf{l}^2+E_0(0)$$  is the center-of-mass part describing a free rotor and
$\mathcal{H}_{r}=\frac{1}{2\Omega}(c_2^a+c_2^b-2c_2^{ab})
\mathbf{l}_r^2+E_0(\theta)$   is  the relative part. Note that
$\cos{\theta}=1-\textbf{n}_r^2/2$ and  $\mathcal{H}_{c}$ and  $\mathcal{H}_{r}$ are decoupled.

Let us focus on the relative motion  and consider   small oscillations around the minimum
$\theta=0$. To the lowest order, we can write  $\textbf{n}_r=q_x\textbf{e}_x+q_y\textbf{e}_y$
and $\textbf{l}_r=l_{rx}\textbf{e}_x+l_{ry}\textbf{e}_y$, with $[q_ i,
l_{rj}]=i\epsilon_{ij}$ ($i,j=x,y$). Then up to a constant,
\begin{equation} \label{relaive}
\mathcal{H}_{r}= \sum_{i=x,y} [\frac{c_2^a+c_2^b-2c_2^{ab}}{2\Omega}
l_{ri}^2+\frac{K}{2}q_i^2].
\end{equation}
where $K=-\frac{\partial E_0}{\partial \cos\theta}|_{\theta=0}$.  Hence  $\mathcal{H}_{r}$ describes
two independent and identical harmonic oscillators, both with  frequency
$$\omega_0\equiv \sqrt{\frac{(c_2^a+c_2^b-2c_2^{ab})K}{\Omega}}=
\sqrt{(\frac{g_a}{N_a}+\frac{g_b}{N_b}-2\frac{g_{ab}}{\sqrt{N_aN_b}})K}.$$
For typical values $g_a\sim g_b\sim 100$ Hz (in the unit of $\hbar$), $N_a\sim N_b\sim 10^4$,
and $g_{ab}/g=0.6$, $K$ can be numerically estimated as $\sim 6g_a$. The corresponding
frequency of the oscillation about the locked position is about $2$Hz.  It can be substantially enhanced, even by a few orders of magnitude, when an optical
lattice is applied and the amplitude of fluctuation is tuned~\cite{song08}. More investigations are needed to address this circumstance.

The oscillation of the spin directors results in the oscillation of occupation numbers in the
Zeeman sublevels. As an example, let  us consider
the case $\mathbf{l}=\mathbf{l}_a+\mathbf{l}_b=0$ so that $ \mathbf{n}$ is independent of time.
Then $$\textbf{n}_{a}=\mathbf{n}+
\frac{c_2^a-c_2^{ab}}{c_2^a+c_2^b-2c_2^{ab}}\textbf{n}_r$$
and $$\textbf{n}_{b}=\mathbf{n}-
\frac{c_2^b-c_2^{ab}}{c_2^a+c_2^b-2c_2^{ab}}\textbf{n}_r. $$  Thus the spin states  of
species $a$ and $b$ in the Zeeman basis state of $m=\pm 1$ are $$\xi_{a \pm 1}=
\frac{c_2^a-c_2^{ab}}{\sqrt{2}(c_2^a+c_2^b-2c_2^{ab})}(iq_y\mp q_x)$$  and  $$\xi_{b \pm 1}=-
\frac{c_2^b-c_2^{ab}}{\sqrt{2}(c_2^a+c_2^b-2c_2^{ab})}(iq_y\mp q_x), $$ while the spin states
of species $a$ and $b$ in the Zeeman basis state of $m=0$ are both $\mathbf{n}$.  Since $q_x$
and $q_y$ both oscillate with frequency $\omega_0$, the
occupation number $N_\alpha |\xi_{\alpha \pm 1}|^2$   oscillates with  frequency $2\omega_0$. The occupation numbers may be probed by  using,  say,  an optical cavity~\cite{cui}.

\section{Locking in a trap  \label{trap}}

Now we examine how the locking effect  survives  a trapping potential, which is an
experimental necessity. Supposing that the potential has the harmonic form  and the   clouds of the two species   have the
same size $R$, then we have
\begin{equation}\label{dense}
    \rho_\alpha=A_\alpha (R^2-r^2),
\end{equation}
where $A_\alpha$ is a positive constant.

The spin dynamics is  determined by the Heisenberg equations $$i\partial_t
\textbf{n}_\alpha(\textbf{r})=[\textbf{n}_\alpha(\textbf{r}), \mathcal{H}_s], $$ $$i\partial_t
\textbf{L}_\alpha(\textbf{r})=[\textbf{L}_\alpha(\textbf{r}), \mathcal{H}_s].$$ For small
fluctuations, we can impose the commutation relations $$
[\mathbf{L}^i_\alpha(\mathbf{r}), \mathbf{n}^j_\beta(\mathbf{r}')]=
i\delta_{\alpha\beta}\epsilon^{ijk}\mathbf{n}^k_\alpha(\mathbf{r})
\delta(\mathbf{r}-\mathbf{r}'), $$
$$[\mathbf{L}^i_\alpha(\mathbf{r}), \mathbf{L}^j_\beta(\mathbf{r}')]
=  i\delta_{\alpha\beta}\epsilon^{ijk}\mathbf{L}^k_\alpha(\mathbf{r})
\delta(\mathbf{r}-\mathbf{r}'),$$
where $i, j, k=x, y, z$. One obtains
\begin{equation}
\begin{array}{rcl}
{\partial}_t\mathbf{n}_\alpha & = & c_2^\alpha(\mathbf{n}_\alpha \times\mathbf{L}_\alpha)
+c_2^{ab}(\mathbf{n}_\alpha\times\mathbf{L}_{\beta}), \\
{{\partial}}_t\mathbf{L}_\alpha & = & \frac{1}{m_\alpha}
\mathbf{n}_\alpha\times\nabla \cdot (\rho_\alpha\nabla\mathbf{n}_\alpha)  +
c_2^{ab} \mathbf{L}_\beta\times\mathbf{L}_\alpha,
\end{array} \label{spin}
\end{equation}
where $\alpha\neq\beta$ and  $\nabla^2 \sqrt{\rho}$ terms are neglected for large clouds.   In
the Thomas-Fermi approximation, the spin structure for the ground state remains unaffected, thus
the fluctuating  directors can be generally written as  $$\textbf{n}_a= u_{ax}
\textbf{e}_x+u_{ay}\textbf{e}_y+ \sqrt{1-u_{ax}^2-u_{ay}^2}  \textbf{n}_a^0$$   and
$$\textbf{n}_b=u_{bx} \textbf{e}_x'+
u_{by'}\textbf{e}_y+\sqrt{1-u_{bx}^2-u_{by}^2}\textbf{n}_b^0.$$ To   first order of $u_{\alpha
i}$ and $L_{\alpha i}$,
\begin{equation}\label{en}
\begin{array}{rcl}
\partial_t u_{\alpha y}&=&c_2^\alpha L_{ \alpha x}+c_2^{ab}L_{\beta x}\cos \theta, \\
\partial_t L_{\alpha x}&=&\frac{1}{m_\alpha}\nabla\cdot(\rho_\alpha \nabla  u_{\alpha y}).  \\
\end{array}
\end{equation}
The two eigenfrequencies  are given by
\begin{equation}
 \omega_{\pm}^2=\frac{f_{nl}}{2}\big[
 \nu_a^2+\nu_b^2\pm[(\nu_a^2-\nu_b^2)^2
 +4\nu_{ab}^2\nu_{ba}^2\cos^2\theta]^{1/2}
 \big],\end{equation}
where $f_{nl}\equiv 2n^2+2nl+3n+l$, $l$ is the angular quantum number, $2n$  is the order of a polynomial of even powers describing the radial wave function~\cite{stringari}, $\nu_a^2=\frac{Ac_2^a}{m_a}$,
$\nu_b^2=\frac{Bc_2^b}{m_b}$, $\nu_{ab}^2=\frac{Bc_2^{ab}}{m_b}$, and
$\nu_{ba}^2=\frac{Ac_2^{ab}}{m_a}$.

Therefore
$\omega_{+}+\omega_{-}=(f_{nl})^{1/2}
   [\nu_a^2+\nu_b^2+2(\nu_a^2\nu_b^2-\nu_{ab}^2\nu_{ba}^2\cos^2\theta
   )^{1/2}]^{1/2}$,
which reaches its minimum at $\theta=0$. Hence the  locking also occurs in a trap.

Now let us address the effect of an external magnetic field  along  the $z$ direction, the
presence of which breaks the full rotational symmetry   down  to  $S^1$ symmetry,  leading to a
nonzero mean-field value of $L_{\alpha z}$. It can be shown that in the mean-field
ground state~\cite{stenger}, each  species $\alpha$ undergoes Bose-Einstein condensation with
the Zeeman basis wave function $(\psi_{\alpha,1}, 0, \psi_{\alpha,-1})^T$, hence the ground
state only possesses spin rotation symmetry $S^1\times S^1$, as the two spins can rotate
around the $z$ axis independently without changing the mean-field energy. Due to the Zeeman barrier,
the low energy dynamics is dominated by the phase fluctuations of $\psi_{\alpha,1}$ and
$\psi_{\alpha,-1}$, while $\psi_{\alpha,0}$ remains zero,  thus the effective Hamiltonian
describes a mixture of two pseudospin-$\frac{1}{2}$ Bose gases~\cite{shi1}.  For low enough magnetic fields, the
spin fluctuation can overcome the Zeeman barrier and restore  $S^2$ symmetry~\cite{zhou1}, consequently the locking persists.

\section{Summary \label{summ}}

Note that the spin directors are macroscopic collective variables of the spinor Bose gases, due
to Bose-Einstein condensation. We have shown that in our system, microscopic quantum fluctuations dramatically change the nature of spontaneous symmetry breaking.
Two interacting macroscopic systems undergo spontaneous symmetry breaking in a correlated way, and consequently the  two macroscopic collective variables are locked.
The symmetry-breaking states of the system are
$|N_a,\mathbf{n}\rangle \otimes |N_b,\mathbf{n}\rangle$,
where the spin directors of the two species are locked to be ${\bf n}$  along  an arbitrary direction.
They are in contrast to states $|N_a,\mathbf{n}_a\rangle \otimes |N_b,\mathbf{n}_b\rangle$, as suggested by the simple mean-field analysis, where $\mathbf{n}_a$ and $\mathbf{n}_b$ are arbitrary and independent of each other.

To summarize, by considering  a  mixture of two distinct species of spin-1 atoms with
interspecies spin exchange, we have shown that the zero-point quantum fluctuations lift the ground
state degeneracy suggested by the mean field theory and lead to the locking between the spin directors of  the two species under the experimentally realistic conditions.
This is a  type of quantum phenomenon in which the microscopic  quantum fluctuations fundamentally control the macroscopic collective phenomenon,  by changing the very nature of symmetry breaking.

\acknowledgments

This work was supported by the National Science Foundation of China (Grant No.  11374060) and
NSERC (Canada). F.Z. was also supported by Canadian Institute for Advanced Research.

\end{document}